\documentclass[reprint,aps,prd,twocolumn,superscriptaddress]{revtex4-2}
\usepackage{amsmath,amssymb}
\usepackage{graphicx}

\usepackage[colorlinks=true, 
            linkcolor=blue, 
            citecolor=blue, 
            urlcolor=blue]{hyperref}

\usepackage{orcidlink}

\begin{document}

\title{Limits on Primordial Black Hole Evaporation from LUX-ZEPLIN}

\author{Jitumani Kalita \orcidlink{0009-0001-5681-6194}}
\affiliation{Department of Physics, Indian Institute of Technology Guwahati,\\
Assam, India}
\email{k.jitumani@iitg.ac.in}

\begin{abstract}
The LUX-ZEPLIN (LZ) collaboration recently reported a singular anomalous nuclear recoil event at $248 \pm 23 \, \rm keV$, accompanied by a strict null result for unexplained excesses in their primary $5.4-50 \, \rm keV$ search window. The macroscopic momentum transfer required to generate this event is kinematically inaccessible to standard halo cold dark matter (CDM), naturally motivating models featuring light, relativistic relics. In this work, we investigate whether a contemporary flux of Hawking-boosted dark matter (DM) emitted by evaporating Primordial Black Holes (PBHs) in the $10^{11}-10^{13} \, \rm g $ mass range can source the anomaly. By parameterizing the nuclear scattering with heavy non-relativistic effective field theory (NREFT) operators and endothermic inelastic mass transitions, we formulate a strict kinematic exclusion. We demonstrate that the inherent thermal nature of the Hawking emission, coupled with the relativistic kinematics of the incident flux, inevitably overproduces low-energy recoils, aggressively violating the LZ background bounds. Utilizing this failure, we map the LZ low-energy null results into novel, highly stringent upper limits on the PBH abundance fraction, excluding $f_{\rm PBH} \gtrsim 10^{-6}$ for $M_{\rm PBH} \sim 2 \times 10^{11} \, \rm g $ at $ 90 \% $ confidence level. 
\end{abstract}

\maketitle

\section{Introduction}
\label{sec:intro}

The existence of dark matter (DM) is firmly established by diverse astrophysical and cosmological observations, spanning from galactic rotation curves to the anisotropies of the cosmic microwave background~\cite{Planck:2018vyc, Rubin:1970zza, Clowe:2006eq, Bertone:2004pz, Aghanim:2018eyx}. For decades, the theoretical landscape has been dominated by the Weakly Interacting Massive Particle (WIMP) paradigm~\cite{Lee:1977ua, Steigman:1984ac, Jungman:1995df, Roszkowski:2017nbc}. Consequently, direct detection experiments---particularly dual-phase liquid xenon time projection chambers (LXe TPCs)---have achieved remarkable sensitivity in probing the WIMP-nucleon scattering cross section \cite{Schumann:2019eaa, Billard:2021uyg}. Recent campaigns by the LUX-ZEPLIN (LZ) \cite{LZ:2022lsv}, XENONnT \cite{XENON:2023cxc}, and PandaX-4T~\cite{Meng:2021zcg} collaborations have placed stringent upper bounds on standard halo DM, pushing the limits of the non-relativistic contact interaction framework~\cite{Aprile:2018dbl, Aalbers:2022dzl, Meng:2021zcg}.

Recently, the LZ collaboration reported the results of a search extending into a higher nuclear recoil energy window~\cite{Akerib:2026LZ}. While observing no unexplained excesses in their primary $5.4-50$~keV WIMP search region, they recorded a single anomalous event at $E_R = 248 \pm 23 \; \rm keV$~\cite{Akerib:2026LZ}. This region of the $(S1c, S2c)$ parameter space is highly suppressed for standard backgrounds, resulting in a $3.4\sigma$ local significance~\cite{Akerib:2026LZ}. Explaining such a macroscopic momentum transfer ($q \simeq 246 \; \rm MeV / c $) within the standard cold dark matter (CDM) paradigm is kinematically hostile. Halo WIMPs, constrained by the galactic escape velocity ($v_{\rm esc} \sim 10^{-3} c$), require extremely heavy masses ($m_\chi \gtrsim 74  \; \rm GeV /c^2$) to trigger this recoil, but models fitting this criteria are largely ruled out by the null low-energy results unless heavily modified by inelastic kinematics~\cite{TuckerSmith:2001hy, Cui:2009xq, ArkaniHamed:2008qn, Chang:2008gd, Bramante:2016rdh, Barello:2014uda, De:2026win, Palmisano:2026kuj, Bose:2026ndd, Bose:2026szs, Mahapatra:2026glu, Das:2026uyy, Chattaraj:2026fxn, Chanda:2026wdl, Bandyopadhyay:2026gjw, Borah:2026zwf, Borah:2026ris, Barman:2026omh, Cabo-Almeida:2026uqw, An:2026pkc} or strongly momentum-dependent elastic interactions, such as scattering via a pseudoscalar mediator~\cite{Unwin:2026rdp}.

An alternative pathway to bypass the halo velocity limit is to consider light DM particles boosted to relativistic speeds. Fast-moving light relics can easily carry the required momentum to excite high-energy nuclear recoils \cite{Agrawal:2014ufa}. Recent literature has extensively explored boosted DM mechanisms, including cosmic-ray upscattering \cite{Bringmann:2018cvk, Ema:2018bes, Cappiello:2018hsu, Dent:2019krz, Wang:2023wrx, Chauhan:2026udz}, solar reflection \cite{An:2017ojc, Emken:2021lgc}, and meson decays \cite{Arguelles:2022fqq}. However, these models frequently struggle with stringent low-energy constraints.

A particularly compelling source for a relativistic dark sector flux is the Hawking evaporation of Primordial Black Holes (PBHs)~\cite{Hawking:1974rv, Hawking:1975vcx, Carr:1974nx, Carr:1975qj, Kalita:2025foa, Chatterjee:2025wnt}. Formed from the collapse of large density perturbations in the early radiation-dominated universe~\cite{Sasaki:2018dmp, Carr:2020gox, Green:2020jor, Villanueva-Domingo:2021spv, Escriva:2022duf, Khlopov:2008qy}, PBHs in the ``asteroid mass'' window ($10^{11} \lesssim M_{\rm PBH} \lesssim 10^{13} \; \rm g$) remain a viable candidate for a substantial fraction of the total DM abundance~\cite{Katz:2018zrn, Montero-Camacho:2019jte, Smyth:2019whb, Dasgupta:2019cae, Laha:2019ssq, Das:2025vts, Kalita:2025fcs}. As these PBHs evaporate, they act as a universal source for all particle species lighter than their Hawking temperature, injecting highly relativistic DM particles into the present-day universe \cite{MacGibbon:1991tj, MacGibbon:1990zk, Hooper:2019gtx, PerezGonzalez:2020vnz, Cheek:2021odj, Auffinger:2020afu, Bernal:2020bjf, Masina:2020xhk, Morrison:2018xla, Gondolo:2020uqv}.

In this work, we investigate whether this contemporary flux of Hawking-boosted DM can explain the LZ $248 \; \rm keV$ anomaly. By parameterizing the nuclear scattering with heavy non-relativistic effective field theory (NREFT) operators \cite{Fitzpatrick:2012ix, Anand:2013yka, Gresham:2014vja} and endothermic inelastic mass transitions, we formulate a strict No-Go theorem. We demonstrate that the inherent thermal nature of the Hawking emission inevitably overproduces low-energy recoils in the LZ detector, violating the experimental bounds. Consequently, we utilize this null result to derive stringent, novel upper limits on the allowed PBH DM fraction $f_{\rm PBH}$ in the asteroid-mass regime.

The paper is organized as follows. In Sec.~\ref{sec:kinematics}, we establish the strict kinematic requirements imposed by the LZ $248 \; \rm keV$ anomaly. Sec.~\ref{sec:nogo} calculates the contemporary Hawking-boosted DM flux and demonstrates the failure of both momentum-dependent elastic scattering and endothermic inelastic transitions to suppress the inevitable low-energy recoil excess. In Sec.~\ref{sec:constraints}, we invert this spectral failure to derive new, stringent upper limits on the PBH abundance fraction $f_{\rm PBH}$. Finally, Sec.~\ref{sec:conclusion} provides our concluding remarks.

\section{The Kinematics of the LZ Anomaly}
\label{sec:kinematics}

The recent results from the LUX-ZEPLIN (LZ) experiment, utilizing an extended nuclear recoil energy window over a $2.84$ tonne-year exposure, revealed a single anomalous event characterized by a nuclear recoil energy of $E_R = 248 \pm 23  \; \rm keV$. This event resides in a region of the $(S1c, S2c)$ parameter space where the expected background is highly suppressed, resulting in a $3.4\sigma$ local significance for models favoring high-energy recoils. To assess the viability of any DM interpretation, we must first establish the rigid kinematic constraints dictated by this specific energy deposition on a xenon target ($m_N \simeq 122.3 \; \rm GeV /c^2$).

For an incident DM particle $\chi$ of mass $m_\chi$ and total energy $E_\chi$ scattering elastically off a stationary nucleus, the momentum transfer is given by $q = \sqrt{2 m_N E_R}$. To generate the LZ event, the required momentum transfer is macroscopic on the nuclear scale
\begin{equation}
    q \simeq 246 \text{ MeV}/c \implies q \simeq 1.25 \text{ fm}^{-1}.
\end{equation}
This momentum transfer sits squarely on the second diffraction lobe of the Helm form factor for xenon, leading to substantial coherence loss, which any successful theoretical model must counteract.

The maximum recoil energy transferred in a head-on collision ($\theta^* = \pi$ in the center-of-momentum frame) is~\cite{Lewin:1995rx}
\begin{equation}
    E_R^{\max} = \frac{2 m_N (E_\chi^2 - m_\chi^2)}{(m_\chi + m_N)^2 + 2 m_N (E_\chi - m_\chi)}.
\end{equation}
For standard CDM in the galactic halo, the DM velocity is bounded by the galactic escape velocity ($v_{\text{esc}} \sim 544 \; \rm km /s$, or $v/c \sim 1.8 \times 10^{-3}$)~\cite{Smith:2006ym, Deason:2019hsm}. In the non-relativistic limit ($E_\chi \simeq m_\chi$), generating a $248 \; \rm keV$ recoil strictly requires $m_\chi \gtrsim 74 \; \rm Gev /c^2$. However, if we consider a light DM particle ($m_\chi \ll E_\chi \ll m_N$) emitted as a relativistic flux---such as from the Hawking radiation of a PBH---the kinematic threshold simplifies to
\begin{equation}
    E_\chi^{\min} \simeq \sqrt{\frac{m_N E_R}{2}}.
\end{equation}
Evaluating this for the LZ anomaly yields a stringent minimum incident energy
\begin{equation}
    E_\chi^{\min} \simeq 123.1 \text{ MeV}.
\end{equation}
Consequently, any relativistic light relic flux must possess a significant population of particles above $\sim 123 \; \rm MeV$ to dynamically trigger the LZ observation. 

While achieving this $E_\chi^{\min}$ threshold is a necessary condition, it is not sufficient. The LZ collaboration simultaneously reported a null result for unexplained excesses in their primary WIMP search region at lower energies. Specifically, the detection efficiency exceeds $50\%$ above $5.4 \; \rm keV$, yet no anomalous event clusters were observed in the $5.4-50 \; \rm keV$ window.

This introduces a severe spectral tension: a successful model must not only possess the kinematic reach to populate the $248 \; \rm keV$ bin but must also feature a differential scattering cross section $d\sigma/dE_R$ that strictly suppresses the event rate at lower recoil energies. For relativistic fluxes driven by thermal distributions (such as Hawking radiation), particles with $E_\chi \sim 20-50 \; \rm MeV$ are exponentially more abundant than those at $123 \; \rm MeV$. As we will demonstrate in the following section, this interplay between the incident thermal flux and the required kinematic threshold establishes a strict No-Go theorem for elastic scattering of boosted relics.

\section{Hawking-Boosted Dark Matter}
\label{sec:nogo}

To evaluate if PBH evaporation can generate the LUX-ZEPLIN (LZ) anomaly, we assume that PBHs form in a purely radiation-dominated universe with an equation of state parameter $\omega = 1/3$. The initial PBH mass is taken to be a fraction of the horizon mass at the time of formation, $M_{\rm PBH} = \gamma M_H$, with the collapse fraction $\gamma \simeq 0.2$~\cite{Carr:1975qj, Shibata:1999zs}. For this analysis, we assume a monochromatic PBH mass spectrum. 

As established in Sec.~\ref{sec:kinematics}, the required $123 \; \rm MeV$ kinematic threshold dictates that the PBH must be sufficiently hot. Setting $M_{\rm PBH} = 5 \times 10^{11} \; \rm g$ yields a Hawking temperature of $T_{\rm PBH} \simeq 21.2  \; \rm MeV$. The resulting differential event rate is determined by combining the boosted astrophysical flux with the nuclear cross section
\begin{equation}
    \frac{dR}{dE_R} = \frac{1}{m_N} \int_{E_\chi^{\min}}^\infty dE_\chi \, \frac{d\Phi_\chi}{dE_\chi} \frac{d\sigma}{dE_R}(E_\chi, E_R).
\end{equation}
The total differential flux of DM particles arriving at the detector, $d\Phi_\chi/dE_\chi$, consists of both a Galactic (MW) and an extragalactic (EG) component
\begin{equation}
    \frac{d\Phi_\chi}{dE_\chi} = \frac{d\Phi_\chi^{\rm MW}}{dE_\chi} + \frac{d\Phi_\chi^{\rm EG}}{dE_\chi}.
\end{equation}
The Galactic component dominates the local flux. It is obtained by integrating the PBH density along the line of sight over the Milky Way halo
\begin{equation}
    \frac{d\Phi_\chi^{\rm MW}}{dE_\chi} = \frac{f_{\rm PBH}}{4\pi M_{\rm PBH}} \frac{d^2 N_\chi}{dt\, dE_\chi} D_{\rm MW},
\end{equation}
where $f_{\rm PBH} = \Omega_{\rm PBH}/\Omega_{\rm DM}$ is the PBH abundance fraction, $d^2N_\chi/dt\,dE_\chi$ is the instantaneous Hawking emission rate, and $D_{\rm MW} = \int d\Omega \int ds \, \rho_{\rm MW}(r(s, \theta))$ is the standard astrophysical $D$-factor evaluated using a Navarro-Frenk-White (NFW) DM density profile~\cite{Navarro:1995iw}.

The extragalactic component incorporates the redshifted emission from the cosmological PBH background, integrated over cosmic time~\cite{MacGibbon:1991tj, Boudaud:2018hqb}
\begin{equation}
    \frac{d\Phi_\chi^{\rm EG}}{dE_\chi} = \int_0^{z_{\max}} \frac{c\, dz}{H(z)} \frac{\rho_{\rm DM,0} f_{\rm PBH}}{M_{\rm PBH}} \left[ \frac{d^2 N_\chi}{dt\, dE'_\chi} \right]_{E'_\chi = (1+z)E_\chi},
\end{equation}
where $\rho_{\rm DM,0}$ is the present-day cosmic DM density, $H(z)$ is the Hubble parameter, and the Hawking emission spectrum is evaluated at the blueshifted energy $E'_\chi = (1+z)E_\chi$. For both components, the primary emission rate $d^2N_\chi/dt\,dE_\chi$ is governed by the geometrical optics limit of the Hawking greybody factor, heavily modified by the exponential Boltzmann suppression $\exp(-E_\chi / T_{\rm PBH})$ at high energies.

\subsection{Elastic Scattering with Momentum-Dependent Operators}

Standard contact interactions ($d\sigma/dE_R \propto E_\chi^{-2} |F(E_R)|^2$) create a massive overproduction of low-energy recoils because the thermal Hawking flux peaks at lower energies and the Helm form factor $|F(E_R)|^2 \approx 1$ as $E_R \to 0$. To suppress this low-energy excess, the non-relativistic effective field theory (NREFT) framework \cite{Fitzpatrick:2012ix, Anand:2013yka} introduces momentum-dependent operators. 

For a relativistic incident particle ($p_\chi \simeq E_\chi$), we parameterize the matrix element with an explicit momentum transfer dependence $(q/\Lambda)^{2n}$. Factoring in the nuclear coherence, this yields the differential cross section
\begin{equation}
    \frac{d\sigma}{dE_R} = \frac{A^2 \lambda^2}{32 \pi m_N E_\chi^2} \left( \frac{2 m_N E_R}{\Lambda^2} \right)^n |F(E_R)|^2.
\end{equation}
Here, $A$ is the mass number of the target nucleus (for xenon, $A \simeq 131$), which provides the $A^2$ coherent enhancement factor for spin-independent scattering. The parameter $\lambda$ represents the dimensionless effective coupling constant, $\Lambda$ denotes the heavy energy scale of the new physics mediator, and $n$ dictates the momentum dependence stemming from the specific NREFT operator. The function $|F(E_R)|^2$ is the standard Helm nuclear form factor \cite{Helm:1956zz, Lewin:1995rx}, which accounts for the loss of coherence at momentum transfers $q \gtrsim R^{-1}$, where $R$ is the effective nuclear radius.

The LZ collaboration highlighted the magnetic dipole interaction $\mathcal{L}_{10}^s$ \cite{Akerib:2026LZ}, which roughly scales as $n=1$. However, evaluating the rate equation with $n=1$ reveals that the linear $E_R$ suppression is vastly overwhelmed by the exponential Boltzmann advantage of low-energy emission. Producing a 5~keV recoil only requires $E_\chi \simeq 17.5 \, \rm MeV$ (sitting at the peak of the $T_{\rm PBH} \simeq 21.2 \, \rm MeV$ flux), whereas 248~keV requires $E_\chi \simeq 123.1 \; \rm MeV$, incurring a severe phase-space suppression factor of $\sim \exp(-5.8)$.

\begin{figure}
    \centering
    \includegraphics[width=1\linewidth]{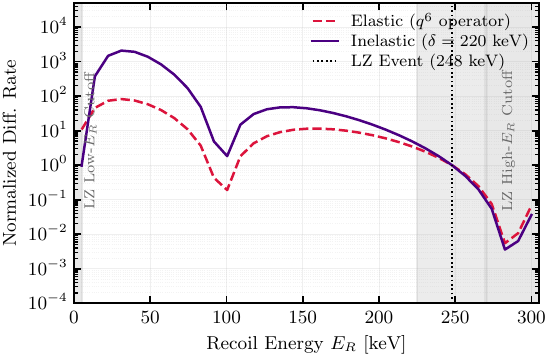}
    \caption{Normalized differential recoil spectra for Hawking-boosted DM scattering on a xenon target. The dashed crimson curve represents elastic scattering suppressed by a heavy momentum-dependent operator ($q^6$ scaling), while the solid indigo curve corresponds to endothermic inelastic scattering with a mass splitting of $\delta = 220 \; \rm keV$. Both theoretical spectra are normalized to the energy of the anomalous LZ event ($248 \; \rm keV$, dotted line). The gray shaded regions designate the energy windows where the LZ detection efficiency drops below $50\%$ \cite{Akerib:2026LZ}. The prominent minima at $\sim 100 \; \rm keV$ and $\sim 280 \; \rm keV$ originate from the diffraction zeroes of the Helm nuclear form factor. Crucially, due to the relativistic nature of the incident Hawking flux, neither extreme momentum dependence nor inelastic kinematics successfully suppresses the primary thermal peak, predicting a massive excess of recoils in the $5.4-50 \; \rm keV$ window where LZ observed a strict null result.}
    \label{fig:recoil_spectra}
\end{figure}

Even pushing to highly suppressed higher-order operators, such as $n=3$ ($q^6$ scaling), fails to rescue the model. As illustrated by the dashed crimson curve in Fig.~\ref{fig:recoil_spectra}, while the extreme momentum dependence forces the rate toward zero at the absolute lowest energies, it introduces a massive secondary peak around $20-30 \; \rm keV$. Numerical integration shows this secondary peak remains roughly two orders of magnitude higher than the predicted rate at $248 \; \rm keV$. Because LZ observed no such excess in the background-free $5.4-50 \; \rm keV$ window, elastic scattering of Hawking-boosted DM is firmly ruled out.

\subsection{Inelastic Scattering of Relativistic Fluxes}

To completely truncate the low-energy spectrum, LZ and contemporary studies rely on endothermic inelastic scattering ($\chi_1 + N \to \chi_2 + N$) with a mass splitting $\delta = m_{\chi_2} - m_{\chi_1} \approx 200-300 \; \rm keV$. A heavy, slow halo WIMP possesses a total kinetic energy ($\sim 30-100 \; \rm keV$) comparable to the mass splitting $\delta$. The kinematic requirement to cross the mass threshold forces the DM to deposit nearly all its momentum into the nucleus, strictly forbidding low-$E_R$ recoils. 

However, we find that this kinematic truncation vanishes for relativistic fluxes. By evaluating the scattering kinematics in the fully relativistic regime \cite{Giudice:2017zhl, Kim:2016zjx}, the minimum incident energy for an inelastic transition becomes
\begin{equation}
    E_\chi^{\min} = \frac{1}{2} \left( \sqrt{2m_N E_R} + (E_R + \delta)\sqrt{\frac{m_N}{2E_R}} \right).
\end{equation}
Because the Hawking-boosted DM is highly relativistic ($E_\chi \gtrsim 123 \, \rm MeV$), the energy cost of the mass splitting ($\delta \sim 220 \, \rm keV$) is negligible, i.e., $E_\chi \gg \delta$. The relativistic particle effortlessly overcomes the transition threshold and is still free to undergo shallow scattering (small angle deflection). Consequently, the incident flux can still easily deposit $E_R \sim 5-50 \, \rm keV$ into the xenon nucleus while transitioning to $\chi_2$. 

This kinematic trap is explicitly demonstrated by the solid indigo curve in Fig.~\ref{fig:recoil_spectra}. Numerical integration of the inelastic cross section over the PBH emission spectrum confirms that the low-energy excess persists, outstripping the $248 \, \rm keV$ event rate by over three orders of magnitude. Thus, inelasticity fails to truncate the low-energy spectrum for relativistic incident fluxes. We conclude a strict No-Go theorem: the LZ $248 \, \rm keV$ anomaly cannot be explained by contemporary relativistic emission from evaporating PBHs, neither through elastic momentum-dependent operators nor through endothermic mass transitions.

\section{Constraints on the Primordial Black Hole Abundance}
\label{sec:constraints}

Having established that Hawking-boosted DM inevitably overproduces low-energy nuclear recoils, we can invert this result to place stringent constraints on the allowed fraction of DM composed of PBHs, $f_{\rm PBH} = \Omega_{\rm PBH} / \Omega_{\rm DM}$. 

The LUX-ZEPLIN collaboration reported a null result for unexplained backgrounds in their primary WIMP search region, demonstrating a detection efficiency exceeding $50\%$ above $5.4 \, \rm keV$~\cite{Akerib:2026LZ}. The extended search window up to $\sim 270 \, \rm keV$ observed only the single anomalous event at $248 \, \rm keV$~\cite{Akerib:2026LZ}. Consequently, the integrated event rate in the lower energy bins ($5.4 \le E_R \le 50 \, \rm keV$) must not exceed the experimental upper limits derived from the $2.84$ tonne-year exposure~\cite{Akerib:2026LZ}. The predicted number of low-energy events in the LZ detector is given by
\begin{equation}
    N_{\rm expected} = \mathcal{E} \int_{5.4 \text{ keV}}^{50 \text{ keV}} dE_R \, \epsilon(E_R) \frac{dR}{dE_R},
\end{equation}
where $\mathcal{E} = 2.84$ tonne-years is the exposure and $\epsilon(E_R)$ is the energy-dependent detection efficiency~\cite{Akerib:2026LZ}. Because the differential rate $dR/dE_R$ is strictly proportional to the PBH fraction $f_{\rm PBH}$ via the Galactic and Extragalactic fluxes, we can express the expected event count as a linear function of the abundance
\begin{equation}
    N_{\rm expected}(f_{\rm PBH}, M_{\rm PBH}) = f_{\rm PBH} \times \mathcal{N}_0(M_{\rm PBH}),
\end{equation}
where $\mathcal{N}_0(M_{\rm PBH})$ is the integrated rate assuming $f_{\rm PBH} = 1$ for a monochromatic PBH mass $M_{\rm PBH}$. By demanding that $N_{\rm expected}$ does not exceed the $90\%$ confidence level upper limit on new physics events in the LZ low-energy signal region, we derive a maximum allowed PBH abundance
\begin{equation}
    f_{\rm PBH}^{\max}(M_{\rm PBH}) = \frac{N_{\rm limit}^{\rm LZ}}{\mathcal{N}_0(M_{\rm PBH})}.
\end{equation}
\begin{figure}[t]
    \centering    
    \includegraphics[width=1\linewidth]{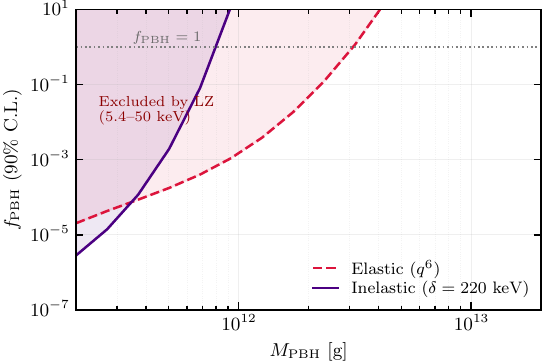}
    \caption{Upper bounds (90\% C.L.) on the PBH abundance fraction $f_{\rm PBH} = \Omega_{\rm PBH}/\Omega_{\rm DM}$ as a function of monochromatic PBH mass $M_{\rm PBH}$, derived from the null results of the LUX-ZEPLIN $2.84$ tonne-year search in the $5.4-50 \; \rm keV$ nuclear recoil window \cite{Akerib:2026LZ}. Constraints are shown for both elastic scattering suppressed by a momentum-dependent operator ($q^6$, dashed crimson) and endothermic inelastic scattering ($\delta = 220 \; \rm keV$, solid indigo). The shaded regions above the respective curves indicate the excluded parameter space. The horizontal dotted line marks $f_{\rm PBH} = 1$. For $M_{\rm PBH} \sim 2 \times 10^{11} \; \rm g$, the elevated Hawking temperature produces an intense relativistic DM flux that yields stringent exclusions down to $f_{\rm PBH} \lesssim 10^{-6}$. For $M_{\rm PBH} \gtrsim 10^{13} \; \rm g$, the lower Hawking temperature drops below the kinematic threshold required to produce $5.4 \; \rm keV$ xenon recoils, resulting in a rapid loss of sensitivity.}    
    \label{fig:fpbh_constraints}
\end{figure}
Applying this limit-setting procedure, we present the $90\%$ C.L. upper bounds on the PBH abundance fraction $f_{\rm PBH}$ in Fig.~\ref{fig:fpbh_constraints}. The constraints are evaluated across the mass range $M_{\rm PBH} \in [2 \times 10^{11}, 2 \times 10^{13}] \; \rm g$ for both the highly suppressed elastic ($q^6$) and endothermic inelastic ($\delta = 220 \; \rm keV$) scattering scenarios. 

As $M_{\rm PBH}$ decreases toward $2 \times 10^{11} \; \rm g$, the PBH becomes hotter ($T_{\rm PBH} \gtrsim 50 \; \rm MeV$), exponentially increasing the emission of relativistic DM particles. This exacerbates the low-energy recoil excess, driving the expected event count $\mathcal{N}_0(M_{\rm PBH})$ higher and consequently strengthening the constraint on $f_{\rm PBH}^{\max}$ down to $\mathcal{O}(10^{-5} - 10^{-6})$. Notably, as shown in Fig.~\ref{fig:fpbh_constraints}, in this lower mass regime the inelastic constraint (solid indigo curve) actually becomes more stringent than the extreme momentum-dependent elastic constraint (dashed crimson curve).

Conversely, for $M_{\rm PBH} \gtrsim 10^{12} \; \rm g$, the Hawking temperature drops substantially. By $M_{\rm PBH} \sim 10^{13} \, \rm g$, the thermal spectrum struggles to populate the high-energy tail required to clear the kinematic threshold for even a minimal $5.4 \; \rm keV$ xenon recoil ($E_\chi^{\min} \simeq 17.5 \; \rm MeV$). This severe Boltzmann suppression rapidly relaxes the constraints, reflected by the sharp upward turn of the exclusion curves in Fig.~\ref{fig:fpbh_constraints} as they cross the physical threshold $f_{\rm PBH} = 1$. Ultimately, this demonstrates that the LZ extended nuclear recoil search provides a highly localized but potent probe of light PBHs, offering complementary constraints to existing astrophysical bounds from Hawking-emitted gamma rays, positrons, X-rays, and other Galactic observables. In particular, complementary constraints on evaporating asteroid-mass PBHs have been derived from Voyager $e^\pm $, inverse-Compton X-rays, and the $511 \; \rm keV$ emission morphology observed by INTEGRAL/SPI~\cite{DelaTorreLuque:2024qms}.


\section{Conclusion}
\label{sec:conclusion}

The observation of a $248 \pm 23 \; \rm keV$ nuclear recoil candidate by the LUX-ZEPLIN experiment \cite{Akerib:2026LZ} necessitates a momentum transfer that cannot be achieved by standard CDM in the galactic halo unless $m_\chi \gtrsim 74 \; \text{GeV}/c^2$. This naturally motivates models featuring light, relativistic relics, such as those generated by the Hawking evaporation of PBHs in the $10^{11}-10^{13} \; \rm g$ mass range. 

In this work, we have demonstrated a strict No-Go theorem for this production mechanism. We parameterized the relativistic scattering utilizing momentum-dependent NREFT operators up to $q^6$ scaling and analyzed the kinematics of endothermic inelastic mass transitions. In all scenarios, the thermal nature of the Hawking emission spectrum intrinsically favors lower-energy particles. When coupled with the rapidly growing xenon nuclear form factor at low momentum transfers, this flux inevitably predicts a massive excess of events in the $5.4-50 \; \rm keV$ energy window, in severe tension with LZ's null results in that regime. 

We conclude that the LZ $248 \; \rm keV$ anomaly cannot be sourced by contemporary Hawking-boosted DM. However, this failure provides a unique phenomenological utility. The rigid absence of a low-energy recoil excess in the LZ data allows us to cast new, stringent upper bounds on the abundance fraction $f_{\rm PBH}$ of light PBHs. Specifically, for PBH masses $M_{\rm PBH} \sim 2 \times 10^{11} \; \rm g$, the resulting constraints exclude PBH DM fractions down to $f_{\rm PBH} \simeq 10^{-6}$ at $90\%$ confidence level. These limits remain highly restrictive across both elastic and inelastic scattering frameworks before naturally relaxing at $M_{\rm PBH} \gtrsim 10^{13} \; \rm g$, where the Hawking temperature drops below the kinematic threshold necessary to excite even a $5.4 \; \rm keV$ xenon recoil. Ultimately, this demonstrates that direct detection experiments, historically designed to probe the non-relativistic galactic halo, double as highly sensitive detectors for the relativistic dark sector of evaporating black holes.

\begin{acknowledgments}
JK would also like to thank Debaprasad Maity and Pankaj Borah for helpful discussions. The work of JK is supported by the Ministry of Human Resource Development, Government of India.
\end{acknowledgments}

\bibliography{References}

@article{Fitzpatrick:2012ix,
    author = "Fitzpatrick, A. Liam and Haxton, Wick and Katz, Emanuel and Lubbers, Nicholas and Xu, Yiming",
    title = "{The Effective Field Theory of Dark Matter Direct Detection}",
    eprint = "1203.3542",
    archivePrefix = "arXiv",
    primaryClass = "hep-ph",
    doi = "10.1088/1475-7516/2013/02/004",
    journal = "JCAP",
    volume = "02",
    pages = "004",
    year = "2013"
}

@article{Anand:2013yka,
  title = {Weakly interacting massive particle-nucleus elastic scattering response},
  author = {Anand, Nikhil and Fitzpatrick, A. Liam and Haxton, W. C.},
  journal = {Phys. Rev. C},
  volume = {89},
  issue = {6},
  pages = {065501},
  numpages = {26},
  year = {2014},
  month = {Jun},
  publisher = {American Physical Society},
  doi = {10.1103/PhysRevC.89.065501}
}

@article{Helm:1956zz,
    author = "Helm, R. H.",
    title = "{Inelastic and Elastic Scattering of 187-Mev Electrons from Selected Even-Even Nuclei}",
    doi = "10.1103/PhysRev.104.1466",
    journal = "Phys. Rev.",
    volume = "104",
    pages = "1466--1475",
    year = "1956"
}

@ARTICLE{Lewin:1995rx,
       author = {{Lewin}, J.~D. and {Smith}, P.~F.},
        title = "{Review of mathematics, numerical factors, and corrections for dark matter experiments based on elastic nuclear recoil}",
      journal = {Astroparticle Physics},
         year = 1996,
        month = dec,
       volume = {6},
       number = {1},
        pages = {87-112},
          doi = {10.1016/S0927-6505(96)00047-3}
}

@unpublished{Akerib:2026LZ,
    author = "Akerib, D. S. and others",
    collaboration = "LZ",
    title = "{Search for dark matter particle interactions in an extended nuclear recoil energy window with the LUX-ZEPLIN (LZ) experiment}",
    eprint = "2609.02823",
    archivePrefix = "arXiv",
    primaryClass = "hep-ex",
    year = "2026"
}

@article{Planck:2018vyc,
    author = "Aghanim, N. and others",
    collaboration = "Planck",
    title = "{Planck 2018 results. VI. Cosmological parameters}",
    eprint = "1807.06209",
    archivePrefix = "arXiv",
    primaryClass = "astro-ph.CO",
    doi = "10.1051/0004-6361/201833910",
    journal = "Astron. Astrophys.",
    volume = "641",
    pages = "A6",
    year = "2020",
    note = "[Erratum: Astron.Astrophys. 652, C4 (2021)]"
}

@article{Rubin:1970zza,
    author = "Rubin, Vera C. and Ford, W. Kent, Jr.",
    title = "{Rotation of the Andromeda Nebula from a Spectroscopic Survey of Emission Regions}",
    doi = "10.1086/150317",
    journal = "Astrophys. J.",
    volume = "159",
    pages = "379--403",
    year = "1970"
}

@article{Clowe:2006eq,
    author = "Clowe, Douglas and Bradac, Marusa and Gonzalez, Anthony H. and Markevitch, Maxim and Randall, Scott W. and Jones, Christine and Zaritsky, Dennis",
    title = "{A direct empirical proof of the existence of dark matter}",
    eprint = "astro-ph/0608407",
    archivePrefix = "arXiv",
    doi = "10.1086/508162",
    journal = "Astrophys. J. Lett.",
    volume = "648",
    pages = "L109--L113",
    year = "2006"
}

@article{Bertone:2004pz,
    author = "Bertone, Gianfranco and Hooper, Dan and Silk, Joseph",
    title = "{Particle dark matter: Evidence, candidates and constraints}",
    eprint = "hep-ph/0404175",
    archivePrefix = "arXiv",
    doi = "10.1016/j.physrep.2004.08.031",
    journal = "Phys. Rept.",
    volume = "405",
    pages = "279--390",
    year = "2005"
}

@article{Aghanim:2018eyx,
    author = "Aghanim, N. and others",
    collaboration = "Planck",
    title = "{Planck 2018 results. I. Overview and the cosmological legacy of Planck}",
    eprint = "1807.06205",
    archivePrefix = "arXiv",
    primaryClass = "astro-ph.CO",
    doi = "10.1051/0004-6361/201833880",
    journal = "Astron. Astrophys.",
    volume = "641",
    pages = "A1",
    year = "2020"
}

@article{Lee:1977ua,
    author = "Lee, Benjamin W. and Weinberg, Steven",
    title = "{Cosmological Lower Bound on Heavy Neutrino Masses}",
    doi = "10.1103/PhysRevLett.39.165",
    journal = "Phys. Rev. Lett.",
    volume = "39",
    pages = "165--168",
    year = "1977"
}

@article{Steigman:1984ac,
    author = "Steigman, G. and Turner, M. S.",
    title = "{Cosmological Constraints on the Properties of Weakly Interacting Massive Particles}",
    doi = "10.1016/0550-3213(85)90537-1",
    journal = "Nucl. Phys. B",
    volume = "253",
    pages = "375--386",
    year = "1985"
}

@article{Jungman:1995df,
    author = "Jungman, Gerard and Kamionkowski, Marc and Griest, Kim",
    title = "{Supersymmetric dark matter}",
    eprint = "hep-ph/9506380",
    archivePrefix = "arXiv",
    doi = "10.1016/0370-1573(95)00058-5",
    journal = "Phys. Rept.",
    volume = "267",
    pages = "195--373",
    year = "1996"
}

@article{Roszkowski:2017nbc,
    author = "Roszkowski, Leszek and Sessolo, Enrico Maria and Trojanowski, Sebastian",
    title = "{WIMP dark matter candidates and searches{\textemdash}current status and future prospects}",
    eprint = "1707.06277",
    archivePrefix = "arXiv",
    primaryClass = "hep-ph",
    reportNumber = "UCI-HEP-TR-2017-09, DO-TH-17-15, UCI-HEP-TR-2017-09-",
    doi = "10.1088/1361-6633/aab913",
    journal = "Rept. Prog. Phys.",
    volume = "81",
    number = "6",
    pages = "066201",
    year = "2018"
}

@article{Schumann:2019eaa,
    author = "Schumann, Marc",
    title = "{Direct Detection of WIMP Dark Matter: Concepts and Status}",
    eprint = "1903.03026",
    archivePrefix = "arXiv",
    primaryClass = "astro-ph.CO",
    doi = "10.1088/1361-6471/ab2ea5",
    journal = "J. Phys. G",
    volume = "46",
    number = "10",
    pages = "103003",
    year = "2019"
}

@article{Billard:2021uyg,
    author = "Billard, J. and others",
    title = "{Direct detection of dark matter\textemdash{}APPEC committee report}",
    eprint = "2104.07634",
    archivePrefix = "arXiv",
    primaryClass = "hep-ex",
    doi = "10.1088/1361-6633/ac5754",
    journal = "Rept. Prog. Phys.",
    volume = "85",
    number = "5",
    pages = "056201",
    year = "2022"
}

@article{LZ:2022lsv,
    author = "Aalbers, J. and others",
    collaboration = "LZ",
    title = "{First Dark Matter Search Results from the LUX-ZEPLIN (LZ) Experiment}",
    eprint = "2207.03764",
    archivePrefix = "arXiv",
    primaryClass = "hep-ex",
    doi = "10.1103/PhysRevLett.131.041002",
    journal = "Phys. Rev. Lett.",
    volume = "131",
    number = "4",
    pages = "041002",
    year = "2023"
}

@article{XENON:2023cxc,
    author = "Aprile, E. and others",
    collaboration = "XENON",
    title = "{First Dark Matter Search with Nuclear Recoils from the XENONnT Experiment}",
    eprint = "2303.14729",
    archivePrefix = "arXiv",
    primaryClass = "hep-ex",
    doi = "10.1103/PhysRevLett.131.041003",
    journal = "Phys. Rev. Lett.",
    volume = "131",
    number = "4",
    pages = "041003",
    year = "2023"
}

@article{Aprile:2018dbl,
    author = "Aprile, E. and others",
    collaboration = "XENON",
    title = "{Dark Matter Search Results from a One Ton-Year Exposure of XENON1T}",
    eprint = "1805.12562",
    archivePrefix = "arXiv",
    primaryClass = "astro-ph.CO",
    doi = "10.1103/PhysRevLett.121.111302",
    journal = "Phys. Rev. Lett.",
    volume = "121",
    number = "11",
    pages = "111302",
    year = "2018"
}

@article{Aalbers:2022dzl,
    author = "Aalbers, J. and others",
    collaboration = "LZ",
    title = "{Projected WIMP sensitivity of the LUX-ZEPLIN dark matter experiment}",
    eprint = "1802.06039",
    archivePrefix = "arXiv",
    primaryClass = "astro-ph.IM",
    doi = "10.1103/PhysRevD.101.052002",
    journal = "Phys. Rev. D",
    volume = "101",
    pages = "052002",
    year = "2020"
}

@article{Meng:2021zcg,
    author = "Meng, Yue and others",
    collaboration = "PandaX-4T",
    title = "{Dark Matter Search Results from the PandaX-4T Commissioning Run}",
    eprint = "2107.13438",
    archivePrefix = "arXiv",
    doi = "10.1103/PhysRevLett.127.261802",
    journal = "Phys. Rev. Lett.",
    volume = "127",
    pages = "261802",
    year = "2021"
}

@article{TuckerSmith:2001hy,
    author = "Tucker-Smith, David and Weiner, Neal",
    title = "{Inelastic dark matter}",
    eprint = "hep-ph/0101138",
    archivePrefix = "arXiv",
    doi = "10.1103/PhysRevD.64.043502",
    journal = "Phys. Rev. D",
    volume = "64",
    pages = "043502",
    year = "2001"
}

@article{Cui:2009xq,
    author = "Cui, Yanou and Morrissey, David E. and Poland, David and Randall, Lisa",
    title = "{Candidates for Inelastic Dark Matter}",
    eprint = "0901.0557",
    archivePrefix = "arXiv",
    primaryClass = "hep-ph",
    doi = "10.1088/1126-6708/2009/05/076",
    journal = "JHEP",
    volume = "05",
    pages = "076",
    year = "2009"
}

@article{ArkaniHamed:2008qn,
    author = "Arkani-Hamed, Nima and Finkbeiner, Douglas P. and Slatyer, Tracy R. and Weiner, Neal",
    title = "{A Theory of Dark Matter}",
    eprint = "0810.0713",
    archivePrefix = "arXiv",
    primaryClass = "hep-ph",
    doi = "10.1103/PhysRevD.79.015014",
    journal = "Phys. Rev. D",
    volume = "79",
    pages = "015014",
    year = "2009"
}

@article{Chang:2008gd,
    author = "Chang, Spencer and Kribs, Graham D. and Tucker-Smith, David and Weiner, Neal",
    title = "{Inelastic Dark Matter in Light of DAMA/LIBRA}",
    eprint = "0807.2250",
    archivePrefix = "arXiv",
    primaryClass = "hep-ph",
    doi = "10.1103/PhysRevD.79.043513",
    journal = "Phys. Rev. D",
    volume = "79",
    pages = "043513",
    year = "2009"
}

@article{Bramante:2016rdh,
    author = "Bramante, Joseph and Fox, Patrick J. and Kribs, Graham D. and Martin, Adam",
    title = "{Inelastic frontier: Discovering dark matter at high recoil energy}",
    eprint = "1608.02662",
    archivePrefix = "arXiv",
    primaryClass = "hep-ph",
    doi = "10.1103/PhysRevD.94.115026",
    journal = "Phys. Rev. D",
    volume = "94",
    number = "11",
    pages = "115026",
    year = "2016"
}

@article{Barello:2014uda,
    author = "Barello, G. and Chang, Spencer and Newby, Christopher N.",
    title = "{A Model Independent Approach to Inelastic Dark Matter Scattering}",
    eprint = "1409.0536",
    archivePrefix = "arXiv",
    primaryClass = "hep-ph",
    doi = "10.1103/PhysRevD.90.094027",
    journal = "Phys. Rev. D",
    volume = "90",
    number = "9",
    pages = "094027",
    year = "2014"
}

@article{Agrawal:2014ufa,
    author = "Agrawal, Prateek and Batell, Brian and Hooper, Dan and Lin, Tongyan",
    title = "{Flavored Dark Matter and the Galactic Center Gamma-Ray Excess}",
    eprint = "1404.1373",
    archivePrefix = "arXiv",
    primaryClass = "hep-ph",
    doi = "10.1103/PhysRevD.90.063512",
    journal = "Phys. Rev. D",
    volume = "90",
    number = "6",
    pages = "063512",
    year = "2014"
}

@article{Bringmann:2018cvk,
    author = "Bringmann, Torsten and Pospelov, Maxim",
    title = "{Novel Direct Detection Constraints on Light Dark Matter}",
    eprint = "1810.10543",
    archivePrefix = "arXiv",
    primaryClass = "hep-ph",
    doi = "10.1103/PhysRevLett.122.171801",
    journal = "Phys. Rev. Lett.",
    volume = "122",
    number = "17",
    pages = "171801",
    year = "2019"
}

@article{Ema:2018bes,
    author = "Ema, Yohei and Sala, Filippo and Sato, Ryosuke",
    title = "{Light Dark Matter at Neutrino Experiments}",
    eprint = "1811.00520",
    archivePrefix = "arXiv",
    primaryClass = "hep-ph",
    doi = "10.1103/PhysRevLett.122.181802",
    journal = "Phys. Rev. Lett.",
    volume = "122",
    number = "18",
    pages = "181802",
    year = "2019"
}

@article{Cappiello:2018hsu,
    author = "Cappiello, Christopher V. and Ng, Kenny C. Y. and Beacom, John F.",
    title = "{Reverse Direct Detection: Cosmic Ray Scattering with Light Dark Matter}",
    eprint = "1810.07705",
    archivePrefix = "arXiv",
    primaryClass = "hep-ph",
    doi = "10.1103/PhysRevD.99.063004",
    journal = "Phys. Rev. D",
    volume = "99",
    number = "6",
    pages = "063004",
    year = "2019"
}

@article{Dent:2019krz,
    author = "Dent, James B. and Dutta, Bhaskar and Newstead, Jayden L. and Shoemaker, Ian M.",
    title = "{Bounds on Cosmic Ray-Boosted Dark Matter in Simplified Models and its Corresponding Observability in Surface-Based Detectors}",
    eprint = "1907.03782",
    archivePrefix = "arXiv",
    primaryClass = "hep-ph",
    doi = "10.1103/PhysRevD.101.116007",
    journal = "Phys. Rev. D",
    volume = "101",
    number = "11",
    pages = "116007",
    year = "2020"
}

@article{Wang:2023wrx,
    author = "Wang, Wenyu and Xu, Wu-Long and Yang, Jin Min and Zhu, Rui",
    title = "{Direct detection of cosmic ray-boosted puffy dark matter}",
    eprint = "2305.12668",
    archivePrefix = "arXiv",
    primaryClass = "hep-ph",
    doi = "10.1016/j.nuclphysb.2023.116348",
    journal = "Nucl. Phys. B",
    volume = "995",
    pages = "116348",
    year = "2023"
}

@article{An:2017ojc,
    author = "An, Haipeng and Pospelov, Maxim and Pradler, Josef and Ritz, Adam",
    title = "{Directly Detecting MeV-scale Dark Matter via Solar Reflection}",
    eprint = "1708.03642",
    archivePrefix = "arXiv",
    primaryClass = "hep-ph",
    doi = "10.1103/PhysRevLett.120.141801",
    journal = "Phys. Rev. Lett.",
    volume = "120",
    number = "14",
    pages = "141801",
    year = "2018"
}

@article{Emken:2021lgc,
    author = "Emken, Tim",
    title = "{Solar reflection of light dark matter with heavy mediators}",
    eprint = "2102.12483",
    archivePrefix = "arXiv",
    primaryClass = "hep-ph",
    doi = "10.1103/PhysRevD.105.055004",
    journal = "Phys. Rev. D",
    volume = "105",
    number = "5",
    pages = "055004",
    year = "2022"
}

@article{Arguelles:2022fqq,
    author = {Arg{\"u}elles, Carlos A. and Mu{\~n}oz, V{\'\i}ctor and Shoemaker, Ian M. and Takhistov, Volodymyr},
    title = "{Hadrophilic light dark matter from the atmosphere}",
    eprint = "2203.12630",
    archivePrefix = "arXiv",
    primaryClass = "hep-ph",
    reportNumber = "IPMU22-0010",
    doi = "10.1016/j.physletb.2022.137363",
    journal = "Phys. Lett. B",
    volume = "833",
    pages = "137363",
    year = "2022"
}

@article{Hawking:1974rv,
    author = "Hawking, S. W.",
    title = "{Black hole explosions}",
    doi = "10.1038/248030a0",
    journal = "Nature",
    volume = "248",
    pages = "30--31",
    year = "1974"
}

@article{Hawking:1975vcx,
    author = "Hawking, S. W.",
    title = "{Particle Creation by Black Holes}",
    doi = "10.1007/BF02345020",
    journal = "Commun. Math. Phys.",
    volume = "43",
    pages = "199--220",
    year = "1975"
}

@article{Carr:1974nx,
    author = "Carr, B. J. and Hawking, S. W.",
    title = "{Black holes in the early Universe}",
    doi = "10.1093/mnras/168.2.399",
    journal = "Mon. Not. Roy. Astron. Soc.",
    volume = "168",
    pages = "399--415",
    year = "1974"
}

@article{Carr:1975qj,
    author = "Carr, B. J.",
    title = "{The Primordial black hole mass spectrum}",
    doi = "10.1086/153853",
    journal = "Astrophys. J.",
    volume = "201",
    pages = "1--19",
    year = "1975"
}

@article{Sasaki:2018dmp,
    author = "Sasaki, Misao and Suyama, Teruaki and Tanaka, Takahiro and Yokoyama, Shuichiro",
    title = "{Primordial black holes\textemdash{}perspectives in gravitational wave astronomy}",
    eprint = "1801.05235",
    archivePrefix = "arXiv",
    primaryClass = "astro-ph.CO",
    doi = "10.1088/1361-6382/aaa7b4",
    journal = "Class. Quant. Grav.",
    volume = "35",
    number = "6",
    pages = "063001",
    year = "2018"
}

@article{Carr:2020gox,
    author = "Carr, Bernard and Kohri, Kazunori and Sendouda, Yuuiti and Yokoyama, Jun'ichi",
    title = "{Constraints on primordial black holes}",
    eprint = "2002.12778",
    archivePrefix = "arXiv",
    primaryClass = "astro-ph.CO",
    reportNumber = "RESCEU-03/20; KEK-Cosmo-249; KEK-TH-2199; IPMU20-0024",
    doi = "10.1088/1361-6633/ac1e31",
    journal = "Rept. Prog. Phys.",
    volume = "84",
    number = "11",
    pages = "116902",
    year = "2021"
}

@article{Green:2020jor,
    author = "Green, Anne M. and Kavanagh, Bradley J.",
    title = "{Primordial Black Holes as a dark matter candidate}",
    eprint = "2007.10722",
    archivePrefix = "arXiv",
    primaryClass = "astro-ph.CO",
    doi = "10.1088/1361-6471/abc534",
    journal = "J. Phys. G",
    volume = "48",
    number = "4",
    pages = "043001",
    year = "2021"
}

@article{Villanueva-Domingo:2021spv,
    author = "Villanueva-Domingo, Pablo and Mena, Olga and Palomares-Ruiz, Sergio",
    title = "{A brief review on primordial black holes as dark matter}",
    eprint = "2103.12087",
    archivePrefix = "arXiv",
    primaryClass = "astro-ph.CO",
    doi = "10.3389/fspas.2021.681084",
    journal = "Front. Astron. Space Sci.",
    volume = "8",
    pages = "87",
    year = "2021"
}

@unpublished{Escriva:2022duf,
    author = "Escriv\`a, Albert and Kuhnel, Florian and Tada, Yuichiro",
    title = "{Primordial Black Holes}",
    eprint = "2211.05767",
    archivePrefix = "arXiv",
    primaryClass = "astro-ph.CO",
    year = "2022"
}

@article{Khlopov:2008qy,
    author = "Khlopov, Maxim Yu.",
    title = "{Primordial Black Holes}",
    eprint = "0801.0116",
    archivePrefix = "arXiv",
    primaryClass = "astro-ph",
    doi = "10.1088/1674-4527/10/6/001",
    journal = "Res. Astron. Astrophys.",
    volume = "10",
    pages = "495--528",
    year = "2010"
}

@article{Katz:2018zrn,
    author = "Katz, Andrey and Kopp, Joachim and Sibiryakov, Sergey and Xue, Wei",
    title = "{Femtolensing by Dark Matter Revisited}",
    eprint = "1807.11495",
    archivePrefix = "arXiv",
    primaryClass = "astro-ph.CO",
    doi = "10.1088/1475-7516/2018/12/005",
    journal = "JCAP",
    volume = "12",
    pages = "005",
    year = "2018"
}

@article{Montero-Camacho:2019jte,
    author = "Montero-Camacho, Paulo and Fang, Xiao and Vasquez, Gabriel and Silva, Makana and Hirata, Christopher M.",
    title = "{Revisiting constraints on asteroid-mass primordial black holes as dark matter candidates}",
    eprint = "1906.05950",
    archivePrefix = "arXiv",
    primaryClass = "astro-ph.CO",
    doi = "10.1088/1475-7516/2019/08/031",
    journal = "JCAP",
    volume = "08",
    pages = "031",
    year = "2019"
}

@article{Smyth:2019whb,
    author = "Smyth, Nolan and Profumo, Stefano and English, Samuel and Jeltema, Tesla and McKinnon, Kevin and Guhathakurta, Puragra",
    title = "{Updated Constraints on Asteroid-Mass Primordial Black Holes as Dark Matter}",
    eprint = "1910.01285",
    archivePrefix = "arXiv",
    primaryClass = "astro-ph.CO",
    doi = "10.1103/PhysRevD.101.063005",
    journal = "Phys. Rev. D",
    volume = "101",
    number = "6",
    pages = "063005",
    year = "2020"
}

@article{Dasgupta:2019cae,
    author = "Dasgupta, Basudeb and Laha, Ranjan and Ray, Anupam",
    title = "{Neutrino and positron constraints on spinning primordial black hole dark matter}",
    eprint = "1912.01014",
    archivePrefix = "arXiv",
    primaryClass = "hep-ph",
    doi = "10.1103/PhysRevLett.125.101101",
    journal = "Phys. Rev. Lett.",
    volume = "125",
    number = "10",
    pages = "101101",
    year = "2020"
}

@article{Laha:2019ssq,
    author = "Laha, Ranjan",
    title = "{Primordial Black Holes as a Dark Matter Candidate Are Severely Constrained by the Galactic Center 511 keV $\gamma$ -Ray Line}",
    eprint = "1906.09994",
    archivePrefix = "arXiv",
    primaryClass = "astro-ph.HE",
    reportNumber = "CERN-TH-2019-099",
    doi = "10.1103/PhysRevLett.123.251101",
    journal = "Phys. Rev. Lett.",
    volume = "123",
    number = "25",
    pages = "251101",
    year = "2019"
}

@article{MacGibbon:1991tj,
    author = "MacGibbon, Jane H.",
    title = "{Quark and gluon jet emission from primordial black holes: The instantaneous spectra}",
    doi = "10.1103/PhysRevD.44.376",
    journal = "Phys. Rev. D",
    volume = "44",
    pages = "376--392",
    year = "1991"
}

@article{MacGibbon:1990zk,
    author = "MacGibbon, J. H. and Webber, B. R.",
    title = "{Quark and gluon jet emission from primordial black holes: The instantaneous spectra}",
    doi = "10.1103/PhysRevD.41.3052",
    journal = "Phys. Rev. D",
    volume = "41",
    pages = "3052--3079",
    year = "1990"
}

@article{Hooper:2019gtx,
    author = "Hooper, Dan and Krnjaic, Gordan and McDermott, Samuel D.",
    title = "{Dark Radiation and Superheavy Dark Matter from Black Hole Domination}",
    eprint = "1905.01301",
    archivePrefix = "arXiv",
    primaryClass = "hep-ph",
    reportNumber = "FERMILAB-PUB-19-186-A",
    doi = "10.1007/JHEP08(2019)001",
    journal = "JHEP",
    volume = "08",
    pages = "001",
    year = "2019"
}

@article{PerezGonzalez:2020vnz,
    author = "P\'erez-Gonz\'alez, Yuber F. and Turner, Jessica",
    title = "{Assessing the tension between a black hole dominated early universe and leptogenesis}",
    eprint = "2010.03565",
    archivePrefix = "arXiv",
    primaryClass = "hep-ph",
    doi = "10.1103/PhysRevD.104.103021",
    journal = "Phys. Rev. D",
    volume = "104",
    number = "10",
    pages = "103021",
    year = "2021"
}

@article{Cheek:2021odj,
    author = "Cheek, Andrew and Heurtier, Lucien and P\'erez-Gonz\'alez, Yuber F. and Turner, Jessica",
    title = "{Primordial black hole evaporation and dark matter production. I. Solely Hawking radiation}",
    eprint = "2107.00013",
    archivePrefix = "arXiv",
    primaryClass = "hep-ph",
    doi = "10.1103/PhysRevD.105.015022",
    journal = "Phys. Rev. D",
    volume = "105",
    number = "1",
    pages = "015022",
    year = "2022"
}

@article{Auffinger:2020afu,
    author = "Arbey, Alexandre and Auffinger, J{\'e}r{\'e}my",
    title = "{BlackHawk: A public code for calculating the Hawking evaporation spectra of any black hole distribution}",
    eprint = "1905.04268",
    archivePrefix = "arXiv",
    primaryClass = "gr-qc",
    reportNumber = "CERN-TH-2019-067",
    doi = "10.1140/epjc/s10052-019-7161-1",
    journal = "Eur. Phys. J. C",
    volume = "79",
    number = "8",
    pages = "693",
    year = "2019"
}

@article{Bernal:2020bjf,
    author = "Bernal, Nicol{\'a}s and Zapata, {\'O}scar",
    title = "{Dark Matter in the Time of Primordial Black Holes}",
    eprint = "2011.12306",
    archivePrefix = "arXiv",
    primaryClass = "astro-ph.CO",
    reportNumber = "PI/UAN-2020-683FT",
    doi = "10.1088/1475-7516/2021/03/015",
    journal = "JCAP",
    volume = "03",
    pages = "015",
    year = "2021"
}

@article{Masina:2020xhk,
    author = "Masina, Isabella",
    title = "{Dark matter and dark radiation from evaporating primordial black holes}",
    eprint = "2004.04740",
    archivePrefix = "arXiv",
    primaryClass = "hep-ph",
    doi = "10.1140/epjp/s13360-020-00564-9",
    journal = "Eur. Phys. J. Plus",
    volume = "135",
    number = "7",
    pages = "552",
    year = "2020"
}

@article{Morrison:2018xla,
    author = "Morrison, Logan and Profumo, Stefano and Yu, Yan",
    title = "{Melanopogenesis: Dark Matter of (almost) any Mass and Baryonic Matter from the Evaporation of Primordial Black Holes weighing a Ton (or less)}",
    eprint = "1812.10606",
    archivePrefix = "arXiv",
    primaryClass = "astro-ph.CO",
    doi = "10.1088/1475-7516/2019/05/005",
    journal = "JCAP",
    volume = "05",
    pages = "005",
    year = "2019"
}

@article{Gondolo:2020uqv,
    author = "Gondolo, Paolo and Sandick, Pearl and Shams Es Haghi, Barmak",
    title = "{Effects of primordial black holes on dark matter models}",
    eprint = "2009.02424",
    archivePrefix = "arXiv",
    primaryClass = "hep-ph",
    doi = "10.1103/PhysRevD.102.095018",
    journal = "Phys. Rev. D",
    volume = "102",
    number = "9",
    pages = "095018",
    year = "2020"
}

@article{Gresham:2014vja,
    author = "Gresham, M. I. and Zurek, K. M.",
    title = "{Effect of nuclear response functions in dark matter direct detection}",
    eprint = "1401.3739",
    archivePrefix = "arXiv",
    primaryClass = "hep-ph",
    doi = "10.1103/PhysRevD.89.123521",
    journal = "Phys. Rev. D",
    volume = "89",
    number = "12",
    pages = "123521",
    year = "2014"
}

@article{Das:2025vts,
    author = "Das, Santabrata and Haque, Md Riajul and Kalita, Jitumani and Karmakar, Rajesh and Maity, Debaprasad",
    title = "{Impact of general relativistic accretion on primordial black holes}",
    eprint = "2505.15419",
    archivePrefix = "arXiv",
    primaryClass = "astro-ph.CO",
    doi = "10.1103/q4ss-c4mp",
    journal = "Phys. Rev. D",
    volume = "112",
    number = "12",
    pages = "123540",
    year = "2025"
}

@article{Kalita:2025fcs,
    author = "Kalita, Jitumani and Maity, Debaprasad",
    title = "{Revisiting PBH accretion, evaporation and their cosmological consequences}",
    eprint = "2512.07291",
    archivePrefix = "arXiv",
    primaryClass = "astro-ph.HE",
    doi = "10.1088/1475-7516/2026/07/085",
    journal = "JCAP",
    volume = "07",
    pages = "085",
    year = "2026"
}

@unpublished{Kalita:2025foa,
    author = "Kalita, Jitumani and Maity, Debaprasad and Chatterjee, Ayan",
    title = "{Black holes in thermal bath live shorter: implications for primordial black holes}",
    eprint = "2501.11925",
    archivePrefix = "arXiv",
    primaryClass = "hep-th",
    month = "1",
    year = "2025"
}

@article{Chatterjee:2025wnt,
    author = "Chatterjee, Ayan and Kalita, Jitumani and Maity, Debaprasad",
    title = "{Evaporation of primordial black holes in a thermal universe: a thermofield dynamics approach}",
    eprint = "2512.07284",
    archivePrefix = "arXiv",
    primaryClass = "hep-th",
    doi = "10.1007/JHEP04(2026)026",
    journal = "JHEP",
    volume = "04",
    pages = "026",
    year = "2026"
}

@article{Smith:2006ym,
    author = {Smith, Martin C. and others},
    collaboration = {RAVE},
    title = "{The RAVE Survey: Constraining the Local Galactic Escape Speed}",
    eprint = {astro-ph/0611671},
    archivePrefix = {arXiv},
    doi = {10.1111/j.1365-2966.2007.11964.x},
    journal = {Mon. Not. Roy. Astron. Soc.},
    volume = {379},
    pages = {755--772},
    year = {2007}
}

@article{Deason:2019hsm,
   title={The local high-velocity tail and the Galactic escape speed},
   volume={485},
   ISSN={1365-2966},
   url={http://dx.doi.org/10.1093/mnras/stz623},
   DOI={10.1093/mnras/stz623},
   number={3},
   journal={Monthly Notices of the Royal Astronomical Society},
   publisher={Oxford University Press (OUP)},
   author={Deason, Alis J and Fattahi, Azadeh and Belokurov, Vasily and Evans, N Wyn and Grand, Robert J J and Marinacci, Federico and Pakmor, Rüdiger},
   year={2019},
   month=Mar, pages={3514–3526} }

@article{Shibata:1999zs,
    author = {Shibata, Masaru and Sasaki, Misao},
    title = "{Black hole formation in the Friedmann universe: Formulation and computation in numerical relativity}",
    eprint = {gr-qc/9905064},
    archivePrefix = {arXiv},
    doi = {10.1103/PhysRevD.60.084002},
    journal = {Phys. Rev. D},
    volume = {60},
    pages = {084002},
    year = {1999}
}

@article{Navarro:1995iw,
    author = {Navarro, Julio F. and Frenk, Carlos S. and White, Simon D. M.},
    title = "{The Structure of cold dark matter halos}",
    eprint = {astro-ph/9508025},
    archivePrefix = {arXiv},
    doi = {10.1086/177173},
    journal = {Astrophys. J.},
    volume = {462},
    pages = {563--575},
    year = {1996}
}

@unpublished{De:2026win,
    author = "De, Bibhabasu",
    title = "{The 248 keV LZ Recoil: A Possible Hint of Non-SM-Like Quark Yukawa Couplings with a Scalar-Portal Dark Matter}",
    eprint = "2609.23096",
    archivePrefix = "arXiv",
    primaryClass = "hep-ph",
    month = "9",
    year = "2026"
}

\end{document}